\newcommand\DIMPY{(C$_7$H$_{10}$N)$_2$CuBr$_4$}
\newcommand\DIMPYdz{(C$_7$D$_{10}$N)$_2$Cu$_{1-z}$Zn$_{z}$Br$_4$}
\newcommand\DIMPYz{(C$_7$H$_{10}$N)$_2$Cu$_{1-z}$Zn$_{z}$Br$_4$}

\documentclass[prl,twocolumn,floatfix,preprintnumbers,amsmath,amssymb,superscriptaddress,longbibliography,showpacs]{revtex4-1}

\usepackage{amsmath,amssymb,bm}
\usepackage{graphicx}
\usepackage{epstopdf}
\usepackage{latexsym}
\usepackage{subfigure}
\usepackage{color}
\usepackage{hyperref}
\usepackage{xspace}
\usepackage[dvipsnames]{xcolor}

\begin{document}

% --------------------------------------------------------------------------------------------------
% Title

\title{Emergent interacting spin islands in a depleted strong-leg Heisenberg ladder}

% --------------------------------------------------------------------------------------------------
% Authors

 \author{D. Schmidiger}
 \email{schmdavi@phys.ethz.ch}
 \affiliation{Neutron Scattering and Magnetism, Laboratory for Solid State Physics, ETH Z\"urich, CH-8093 Z\"urich, Switzerland}

 \author{K. Yu. Povarov}
 \affiliation{Neutron Scattering and Magnetism, Laboratory for Solid State Physics, ETH Z\"urich, CH-8093 Z\"urich, Switzerland}

 \author{S. Galeski}
 \affiliation{Neutron Scattering and Magnetism, Laboratory for Solid State Physics, ETH Z\"urich, CH-8093 Z\"urich, Switzerland}

 \author{N. Reynolds}
 \altaffiliation[Present address:]{ Laboratory for Neutron Scattering, Paul Scherrer Institut, CH-5232 Villigen, Switzerland.}
 \affiliation{Neutron Scattering and Magnetism, Laboratory for Solid State Physics, ETH Z\"urich, CH-8093 Z\"urich, Switzerland}

 \author{R. Bewley}
 \affiliation{ISIS Facility, Rutherford Appleton Laboratory, Chilton, Didcot, Oxon OX11 0QX, United Kingdom}

 \author{T. Guidi}
 \affiliation{ISIS Facility, Rutherford Appleton Laboratory, Chilton, Didcot, Oxon OX11 0QX, United Kingdom}

 \author{J. Ollivier}
 \affiliation{Institut Laue-Langevin, 6 rue Jules Horowitz, 38042 Grenoble, France}

 \author{A. Zheludev}
 \email{zhelud@ethz.ch}
 \homepage{http://www.neutron.ethz.ch/}
 \affiliation{Neutron Scattering and Magnetism, Laboratory for Solid State Physics, ETH Z\"urich, CH-8093 Z\"urich, Switzerland}
\date{\today}

\begin{abstract}
Properties of the
depleted Heisenberg spin ladder material series \DIMPYz~have been studied
by the combination of magnetic measurements and neutron
spectroscopy. Disorder-induced degrees of freedom lead to a specific
magnetic response, described in terms
of emergent strongly interacting ``spin island'' objects. The structure and dynamics of the spin islands is studied by high-resolution inelastic neutron scattering.  This allows to determine their spatial shape and to observe their
mutual interactions, manifested by strong spectral in-gap
contributions.
\end{abstract}


\maketitle

In the solid state, even weak perturbations may lead to qualitatively new physics described in terms of
entirely new emergent degrees of freedom and quasiparticles
\cite{Anderson1972}. One such perturbation, known to open the door to a
variety of novel and competing ground states and rich phase
diagrams, is structural or chemical disorder \cite{DagottoScience}. An exciting 
example of disorder-induced emergent degrees of freedom are the \emph{magnetic} objects that appear upon the introduction
of \emph{non-magnetic} impurities in gapped quantum-disordered antiferromagnets (AFs)
\cite{Mikeska_Doping,Kenzelmann2003p1,Bobroff2009p1, Hase1993p1, *Hase1996p1, *Uchinokura2002p1, Lorenzo2015Np1,Guidi2015p1}. These entities may be understood
as spins released from non-magnetic AF singlets by removing their
partner spins. The short-range correlations in the underlying quantum AF
spread these spin degrees of freedom over extended regions (``spin-islands'') around each impurity site~\cite{Mikeska_Doping}. The size of the spin islands is controlled by the correlation length in the parent system, and may be as large as dozens of nanometers. 
Due to their partial overlap, these spin islands interact.
The original quantum AF thus acts as a ``medium'' that hosts a new magnetic system of 
{\itshape mesoscopic} objects and carries interactions between them. Due to these interactions, the emergent system may have its own unique correlations and dynamics.

The impurity-induced formation of localized $S=1/2$ spin objects has been thoroughly studied in gapped $S=1$
Haldane spin chains
\cite{Glarum1991p1,BatloggPhysicaB,Tedoldi1999p1,Uchiyama1999p1,Masuda2002p4,Kenzelmann2003p1}.
Unfortunately, in these highly one-dimensional chain materials, each impurity completely severs the host system.
The emergent spin islands located at the chain segment ends
merely pair up into isolated dimers \cite{Kenzelmann2003p1}, and have no collective dynamics.
This is in contrast to expectations for  Heisenberg spin
ladders, which are composed of \emph{two} neighboring chains linked
by rung interactions $J_{\perp}$. Within this topology, the
non-magnetic impurities have a low chance to disrupt the continuity
of the system and a finite-size segment typically contains a large
number of mutually interacting islands. To date, the study of non-magnetic
impurities in spin ladders was limited to rather complex materials with prohibitively large energy scales, such as Sr(Cu$_{1-z}$Zn$_z$)$_2$O$_3$~\cite{Azuma1997p1} or
Bi(Cu$_{1-z}$Zn$_z$)$_2$PO$_6$~\cite{Bobroff2009p1}. These cases are difficult to describe theoretically, 
as they involve  additional interactions such as cyclic exchange, frustration, or
3D coupling. What is missing is an experimental study of
emergent and interacting spin islands in a {\itshape clean}
realization of the spin ladder model, such that could be compared to theoretical calculations at the quantitative level.
Fortunately, in recent years, a number of new exceptionally good organic spin ladder
compounds were 
discovered~\cite{LandeeTurnbull_EuJInChem_2013_ReviewMagnets}. Among
them, \DIMPY\
(DIMPY)~\cite{Shapiro2007p1,Schmidiger2012p1}  realizes
the rare case of a strong-leg spin ladder, which is of special
interest in the context of the present study. Its dominant leg interactions imply a
significant correlation length and hence significant interactions
between emerging spin objects, even at low concentration. In the
present Letter we show that DIMPY, diluted with nonmagnetic zinc, is
indeed a perfect prototype for testing this physics. By
combining thermodynamic and high-resolution inelastic neutron
scattering experiments with numerical simulations, we are able to directly study the shape and
interactions of the spin islands, and to quantitatively describe the
impurity-induced {\itshape collective} response in the language of these emergent
objects.

% ================================================================================
% Figure 1
%
\begin{figure}[h!t]
\centering
\includegraphics[width=1\columnwidth]{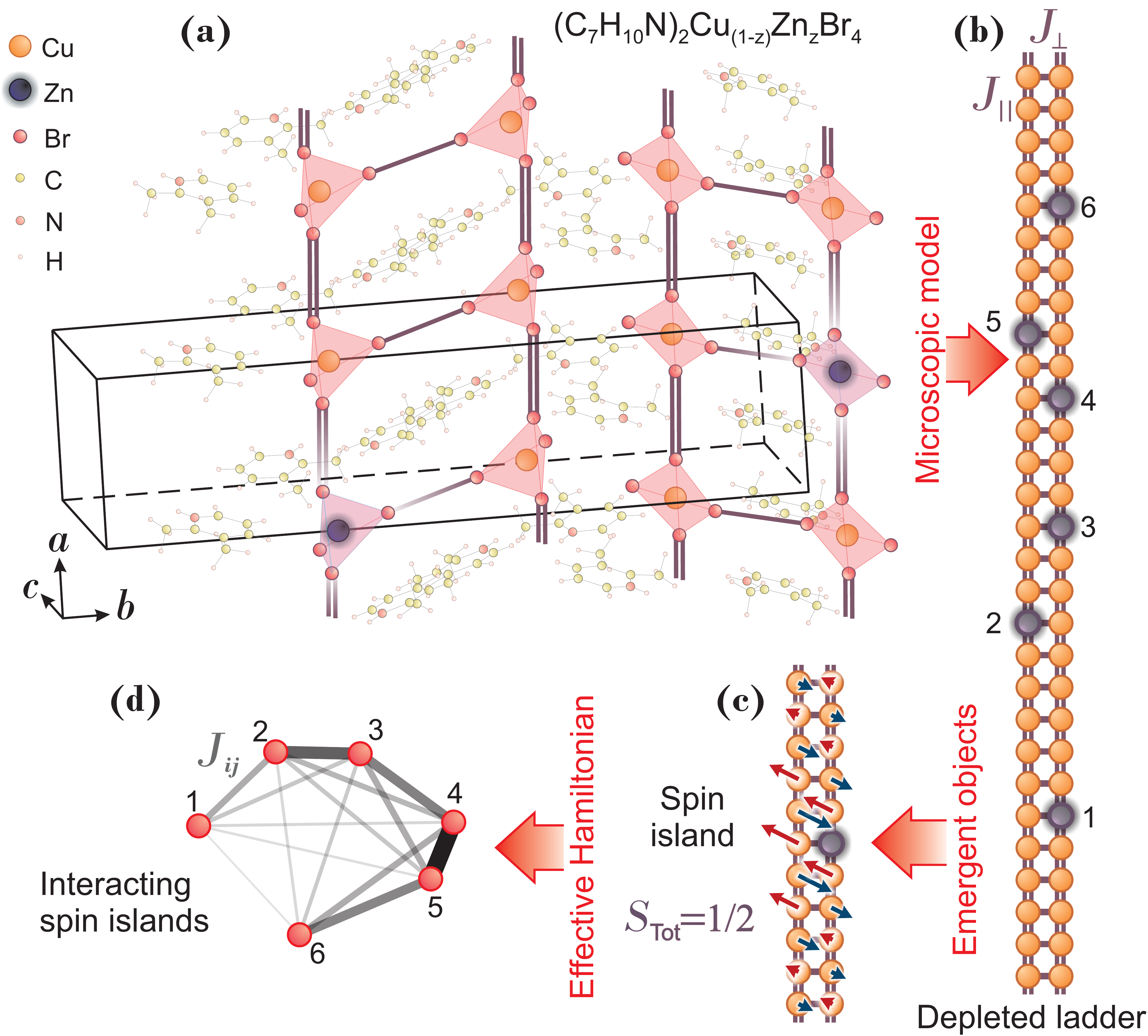}
\caption{From a real material to the microscopic model
and then, to the effective description in terms of emergent objects.
(a) The crystal structure of \DIMPYz. (b) Spin ladder model
microscopic description. The random replacement of magnetic
Cu$^{2+}$ by non-magnetic Zn$^{2+}$ renders some $S=1/2$ sites
missing. (c) Emerging local spin degrees of freedoms (spin islands),
pinned to the impurity position but with a magnetization profile
extending over many unit cells. (d) Effective spin island
Hamiltonian~(\ref{eq:HamEff}) with interactions controlled by the
mutual distances as given by
Eq.~(\ref{eq:Jeff}).}\label{FigSpinIsland}
\end{figure}
%
% ================================================================================

% ================================================================================
% Figure 2
%
\begin{figure}[h!t]
\centering
\includegraphics[width=0.5\textwidth]{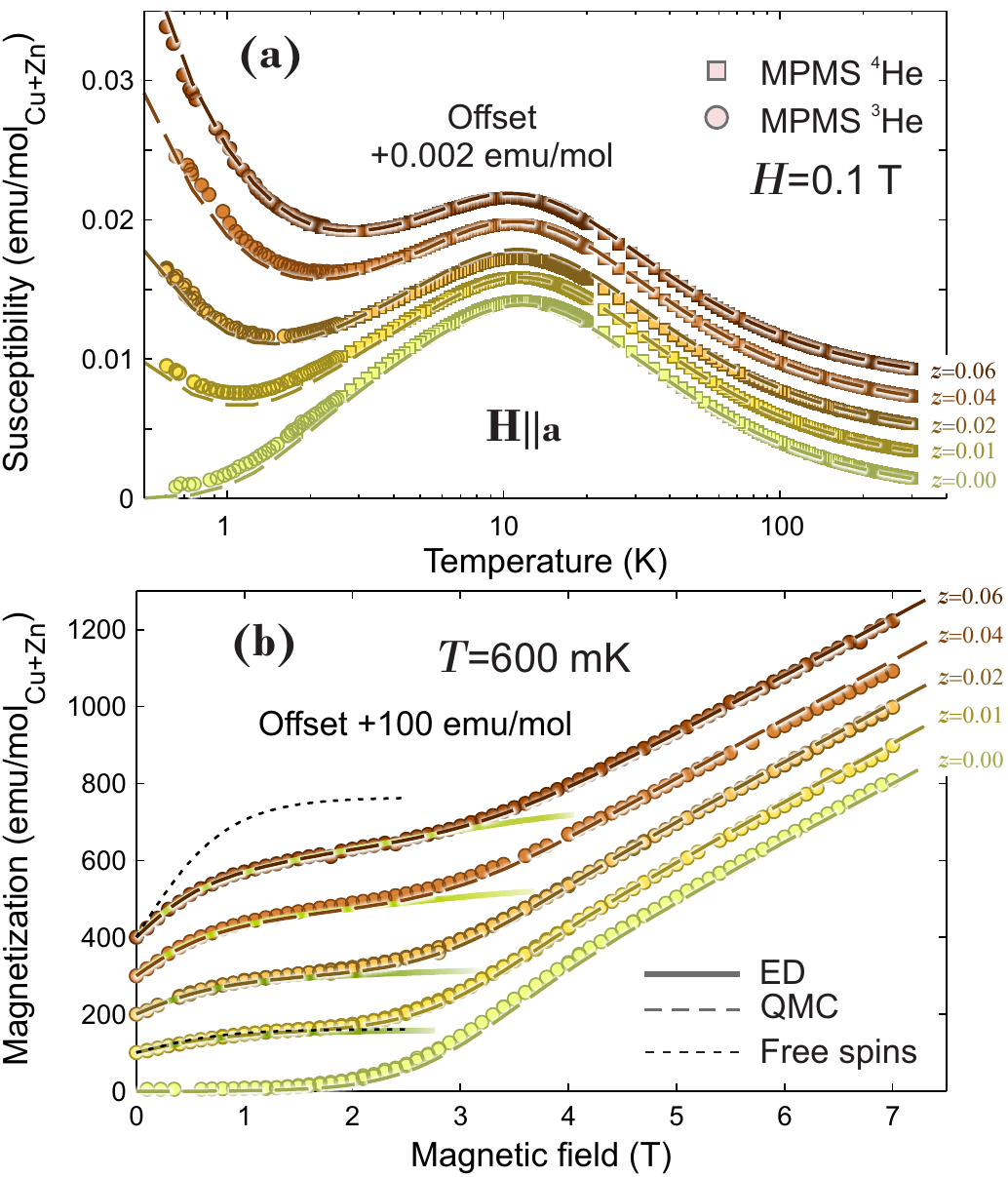}
\caption{Magnetization experiments on the depleted
spin-ladder series (C$_7$H$_{10}$N)$_2$Cu$_{1-z}$Zn$_{z}$Br$_4$. (a)
Susceptibility as a function of temperature. (b) Isothermal
magnetization at $T=600$~mK for the same samples. Symbols are the
experimental data. Numerical QMC simulations and results from the
diagonalization of the effective model [Eq.~(\ref{eq:HamEff})] are
shown as dashed and solid lines correspondingly. Dotted lines in (b)
illustrate the free-spin response for $z=0.01$ and $z=0.06$ cases.}
\label{FigTD}
\end{figure}
% ================================================================================

The parent material DIMPY was shown to be described by a simple
Heisenberg spin-ladder Hamiltonian with $J_{\parallel}=1.42(6)$~meV
along the leg and $J_{\perp}=0.82(2)$~meV (see
Fig.~\ref{FigSpinIsland})~\cite{Schmidiger2012p1,Schmidiger2013p2,PovarovSchmidiger_PRB_2015_DimpyScaling}
and with exceptionally weak additional
interactions~\cite{Jeong2013p1,Glazkov_DIMPYESR,Zvyagin_DIMPYESR}.
We were able to chemically introduce non-magnetic
impurities by replacing a fraction $z$ of magnetic $S=1/2$ Cu$^{2+}$
by non-magnetic Zn$^{2+}$
ions~\cite{Yankova_PhilMag_2012_ReviewXtals}. The
magnetic susceptibility and isothermal magnetization data are
presented in Fig.~\ref{FigTD}, as measured on a series of
(C$_7$H$_{10}$N)$_2$Cu$_{1-z}$Zn$_{z}$Br$_4$ single
crystals with varying impurity concentration $z$~\cite{SupplMat}.
While the susceptibility of the $z=0$ parent compound
is being dominated by a spectral gap of $\Delta=0.33$~meV and
exponentially decays to zero at low temperatures, the susceptibility
of the Zn-diluted material acquires an additional paramagnetic-like
contribution, progressively increasing with the impurity
concentration. This behavior becomes even more
apparent in the low-temperature magnetization
(Fig.~\ref{FigTD}b). The latter remains suppressed by the gap below
the critical field of $H_{c}\simeq 2.7$~T \cite{Schmidiger2012p1} in
the clean material but acquires an impurity-induced
contribution in the Zn-substituted derivatives.
Nonetheless, the observed response is not simply described by the
$S=1/2$ Brillouin function, as it would in case of
free magnetic moments~\cite{Whitebook}. While deviations from the
Brillouin function are small at low impurity concentrations, they
become up to 45\% at $z=0.06$. Qualitatively, this is
explained by the mean impurity distance $\overline{L}_x(z) =
(1-z)/(2z-z^2)$ \cite{Lavarelo2013p1} which is as large as
50 unit cells for $z=0.01$ but becomes comparable to the correlation length
$\xi\approx 6.3$ \cite{SupplMat} at $z=0.06$.
The probability of finding close and strongly
interacting islands thus rapidly increases with impurity concentration.

As a first step, we compare our measurements to numerical
calculations \cite{SupplMat} based on the full parent spin ladder
Hamiltonian~\cite{Schmidiger2012p1}, depleted with randomly placed
non-magnetic sites (Fig.~\ref{FigSpinIsland}b).
Following Ref.~\cite{Lavarelo2013p1}, we calculate the
susceptibility and magnetization with Quantum Monte
Carlo (QMC) simulations of ladder systems with $L=500$ rungs,
$N=2Lz$ randomly placed non-magnetic sites and averaged over $300$
random impurity configurations \cite{ALPSv2p0}. The numeric results (dashed lines in Fig.
\ref{FigTD}) quantitatively reproduce the measured data in the
entire temperature- and field-range. This proves not
only that we are able to chemically control the zinc concentration
but also that its introduction does not lead to
significant local distortion effects altering the
superexchange interactions.

Extensive QMC calculations of large depleted
systems, averaged over hundreds of configurations, are expensive,
time-consuming and yield little physical insight. An alternative,
much simpler approach is provided by the effective low-energy
description in terms of interacting spin islands
(Fig.~\ref{FigSpinIsland}c,d)
\cite{Mikeska_Doping,Lavarelo2013p1}. The latter are represented by
$S=1/2$ spins ${\bf S}_{\bf I}$ at random impurity positions ${\bf
I}$, interacting with distance-dependent effective interactions
$J^\mathrm{eff}({\bf L})$ and described by the Hamiltonian
~\cite{Lavarelo2013p1}
% Eq. 1 =============================================================================
\begin{align}
\mathcal{H}_\mathrm{Sp.Isl.} = \sum_{{\bf I},{\bf J}}
J^\mathrm{eff}({\bf I}-{\bf J}) \; {\bf S}_{{\bf I}} \cdot {\bf
S}_{\bf J} - g\mu_\mathrm{B}H\sum_{\bf I} S_{\bf I}^z,
\label{eq:HamEff}
\end{align}
% ================================================================================
with $\mu_\mathrm{B}$ and $g$ being the Bohr magneton and the
electron $g$-factor. Mutual spin-island interactions are
mediated by the spin correlations in the hosting ladder
medium. Since the latter are short-ranged, effective interactions
exponentially decay with distance. In addition, due to their
antiferromagnetic nature, $J^\mathrm{eff}({\bf L})$ are ferro- or
antiferromagnetic, depending on whether the mutual interaction path
contains an odd or even number of
sites~\cite{Mikeska_Doping,Lavarelo2013p1}, reminiscent of the
oscillatory Ruderman-Kittel-Kasuya-Yosida (RKKY) interaction
\cite{RKKY1,*RKKY2,*RKKY3}. Being fully controlled by the spin
ladder medium, the effective interactions can be numerically
obtained from a system with two non-magnetic sites at distance ${\bf
L}$ in a long ladder system. Following
Refs.~\cite{Lavarelo2013p1,Mikeska_Doping}
we have performed Matrix Product State
calculations~\cite{Dolfi2014p1,ALPSv2p0} and confirmed that
$J^\mathrm{eff}({\bf L})$ is  described by the simple
law
% Eq.2 =============================================================================
\begin{align}
J^\mathrm{eff}({\bf L}) = J_0 (-1)^{L_x+L_y+1}\mathrm{e}^{-|L_x| /
\xi} \label{eq:Jeff}
\end{align}
% ================================================================================
with the energy scale $J_0=0.441$~meV  and the decay length
$\xi=6.28$ for the exchange constants of DIMPY~\cite{SupplMat}. By
including the numerically calculated effective interactions,
Eq.~(\ref{eq:HamEff}) becomes a \emph{parameter-free} description of
the emergent spin island system.

The magnetization and susceptibility of the effective model were
determined \cite{SupplMat} by exactly diagonalizing (ED) the
Hamiltonian in Eq. (\ref{eq:HamEff}), for a system of $N=12$ sites
$S_{\bf I}$, randomly placed on a ladder with $L=N/2z$ rungs and
with the effective interactions given by
Eq.~(\ref{eq:Jeff}). The magnetization, averaged over 5000 random
configurations, is in excellent agreement with measured data (Fig.
\ref{FigTD}b). Notably, a model described by not more than 12
mutually interacting spin operators is sufficient to model the
entire thermodynamic response of this complex disordered many-body
quantum system. Nevertheless, the description in terms of
Eq.~(\ref{eq:HamEff}) remains valid only as long as
the mutual interactions controlled by the spin island correlations
do not change. While the latter remain unaffected by temperature and
applied field as long as the spin ladder remains in its quantum
disordered regime, they fundamentally change once the system crosses
the critical field $H_c \simeq 2.7$~T \cite{Schmidiger2012p1} to the
quantum critical Tomonaga-Luttinger spin liquid (TLSL)
state. Above $H_{c}$ the effective
description naturally fails.

\begin{figure}[t]
\centering
\includegraphics[width=1\columnwidth]{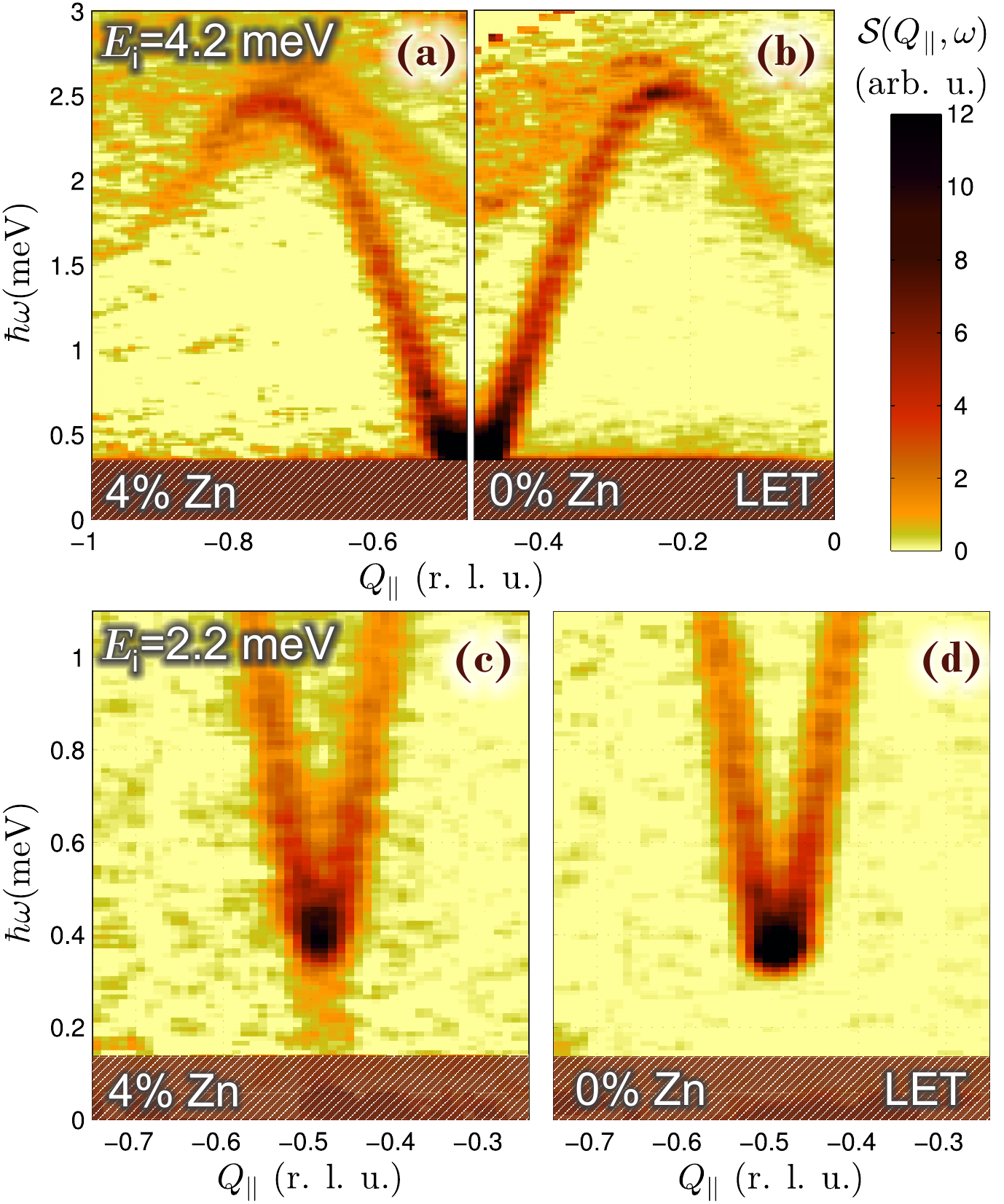}
\caption{Inelastic neutron scattering data on the
depleted and clean spin ladder DIMPY, measured on LET instrument.
(a,b) Entire spectrum in the lower resolution mode (incident energy
$E_{\text{i}} = 4.2$~meV). (c,d) Low-energy sector in the higher
resolution mode ($E_{\text{i}} = 2.2$~meV). Background corrected
magnetic neutron scattering intensity is shown as a function of
energy transfer $\hbar\omega$ and momentum transfer along the leg
$Q_\parallel$. Hatched regions are
contaminated by parasitic nuclear incoherent scattering.}
\label{FigNeutron1}
\end{figure}

So far, we have successfully described the measured thermodynamic
properties in terms of an effective model of emergent spin island.
However, can such objects and their interactions be observed
directly? To answer this question, we present a series of
high-resolution inelastic neutron scattering experiments enabling to
access the spectral properties of both the spin ladder medium and
the low-energy spin island system. In
Fig.~\ref{FigNeutron1}a,c, we show the
background-subtracted magnetic neutron spectrum of
(C$_7$D$_{10}$N)$_2$Cu$_{0.96}$Zn$_{0.04}$Br$_4$, as measured with
the neutron time-of-flight (TOF) technique (LET
spectrometer~\cite{Bewley2011p1}, ISIS facility, U.K.). Data was
gathered in distinct low- and high-resolution setups with incident
energies $E_i = 4.2$~meV and $2.2$~meV ($T=75$~mK). Neutron
intensity is shown as a function of energy transfer $\hbar\omega$
and momentum transfer  $Q_\parallel = {\bf Q}\cdot {\bf a}$ along
the leg direction. Corresponding data from the parent
compound is shown in
Figs.~\ref{FigNeutron1}b,d, measured under
identical experimental
conditions~\cite{PovarovSchmidiger_PRB_2015_DimpyScaling} and
treated in the same way.

% ================================================================================
% Figure 3
%
%
% ================================================================================

First, we observe that the properties of the spin ladder medium
hosting the spin island system remains nearly unaffected by the
presence of impurities (Fig. \ref{FigNeutron1}a,b). In
both the parent and disordered compound, identical
gapped and dispersive magnon and two-magnon bound state branches
\cite{Schmidiger2012p1} are observed. In contrast, clear changes
are found in the low-energy sector (Fig.
\ref{FigNeutron1}c,d). The magnon gap is slightly
shifted from $\Delta=0.33$ to $0.41$~meV, reminiscent
of the magnon blue-shift observed in 1D quantum-disordered
AFs at finite temperature \cite{Zheludev2008p1}. However, as a main 
feature, we observe a stripe of enhanced
in-gap intensity developing around $Q_\parallel = 0.5$~r.l.u. in the
disordered compound. This
contribution originates in strongly interacting spin islands,
leading to a disorder-induced finite density of states inside the
gap.
%
% ================================================================================
% Figure 4
%
\begin{figure}[t]
\centering
\includegraphics[width=1\columnwidth]{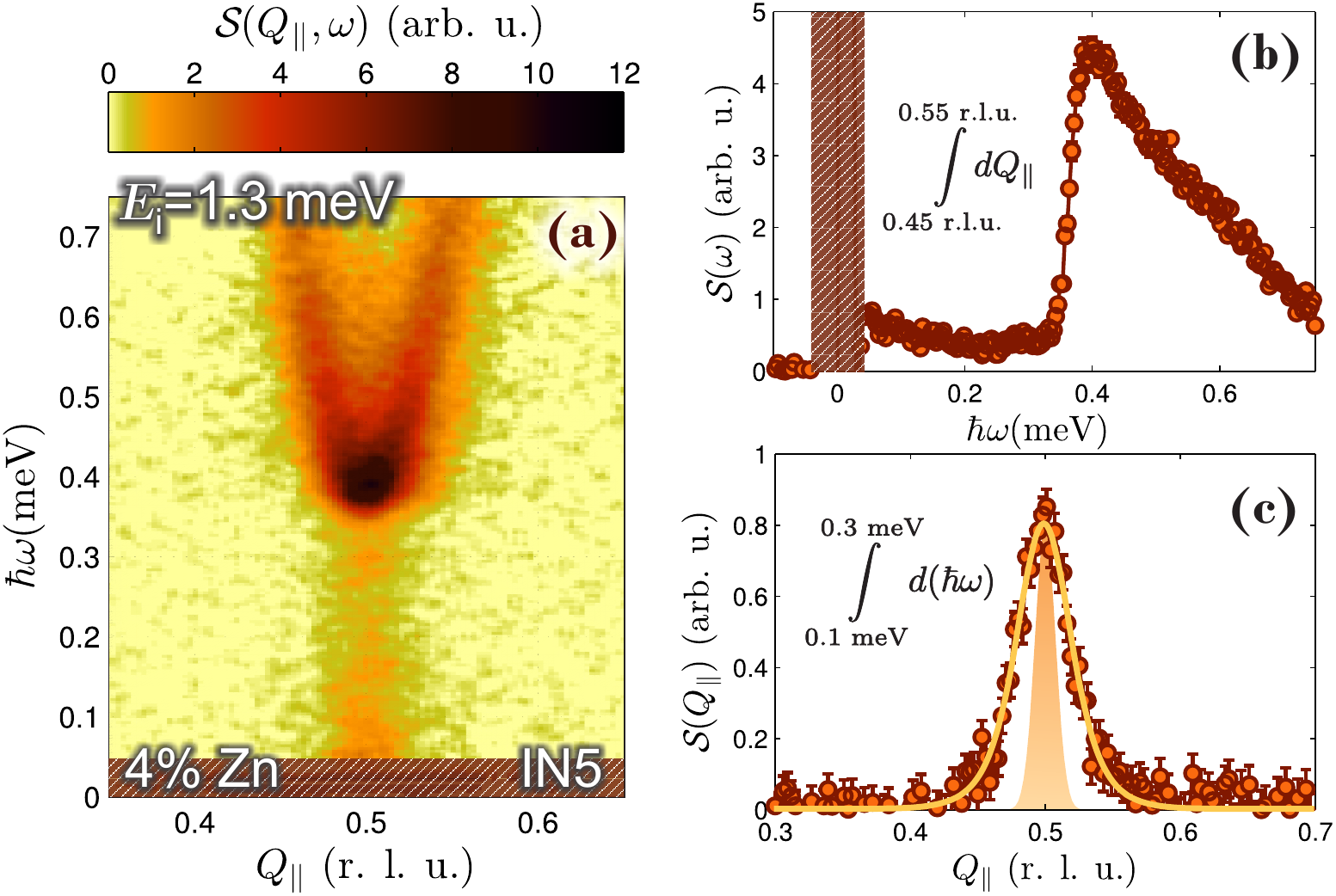}
\caption{Inelastic neutron scattering data on the
disordered DIMPY, measured on IN5 instrument. (a) Background
subtracted spectrum measured with $E_{\text{i}}=1.3$~meV. Neutron
intensity is shown as a function of energy transfer $\hbar\omega$
and momentum transfer $Q_\parallel$. (b) Neutron intensity as a
function of $\hbar\omega$ for $Q_{\parallel}$ between 0.45 and 0.55
r.l.u. Data in the hatched area is contaminated by parasitic nuclear
incoherent scattering. (c) Scattering intensity as a function of
$Q_\parallel$, after integrating in
$0.1\,\mathrm{meV}\le\hbar\omega\le 0.3\,\mathrm{meV}$. Instrumental
resolution (shaded area) is shown along with the
resolution-convoluted fit~(\ref{eq:shape}) [solid line].}
\label{FigNeutron2}
\end{figure}
%
% ================================================================================

It is even more pronounced in the spectrum presented
in Fig.~\ref{FigNeutron2}a, measured with a lower incident energy of
$E_{\text{i}}=1.3$~meV and thus better resolution
(IN5 instrument~\cite{Ollivier2011p1}, ILL, France).
The spin island contribution progressively increases with decreasing energy
$\hbar\omega$, as shown in
Fig.~\ref{FigNeutron2}b. This is readily explained
since at $z=0.04$, the probability of finding spin islands only
weakly interacting with others remains large and there is a
vast number of quantum states at low energies.

As a main characteristics, these in-gap contributions
exhibit an intrinsic $Q_\parallel$ linewidth, as shown
in Fig.~\ref{FigNeutron2}c. This intrinsic linewidth
is a measure of the spin island real-space magnetization profile.
Similar to conventional magnetic neutron
scattering~\cite{squiresbook}, where intensity is modulated by the
magnetic form factor (the Fourier transform of the magnetic ions'
unpaired electron density), scattering by spin islands is modulated
by the square of the spin island form factor
$F(Q_\parallel,Q_\perp)$ --- the Fourier transform of its real space
shape. A similar reasoning was applied to determine
the spatial shape of emergent local objects in the substantially
different cases of the charge-doped or depleted Haldane
spin chains Y$_{2-z}$Ca$_{z}$BaNiO$_5$
\cite{Xu2000p1}, Y$_2$BaNi$_{1-z}$Mg$_z$O$_5$
\cite{Kenzelmann2003p1}. For a spin island with
exponentially decaying staggered magnetization on a ladder, the form
factor reads as
% Eq. 3 ==============================================================================
\begin{align}
F(Q_\parallel,Q_\perp) = \frac{(1-\mathrm{e}^{-\mathrm{i}2\pi Q_\perp})\sinh(\xi^{-1})}{\cos(2\pi Q_\parallel) + \cosh(\xi^{-1})} - 1.
\label{eq:shape}
\end{align}
% ================================================================================
Here, $\xi$ denotes the real space decay length and
$Q_\perp$ is the momentum transfer along the
rung~\cite{Schmidiger2011p1}. We compared
$F|(Q_\parallel,Q_\perp)|^2$, averaged over $Q_\perp$, to our
experimental data and by convolution with experimental resolution,
we were able to quantify the real space decay length to be $\xi =
5.0(2)$. The magnetization profile thus decays to 10\% of its
initial value only after 15 unit cells and the size of an emergent
spin island is of the order of 20 nm. Notably, the determined decay
length is slightly shorter than the numerically calculated value of
$\xi = 6.27$ \cite{SupplMat}. However, this is not surprising since
in-gap states observed in experiments are located at a comparatively
large energy transfer on the order of 0.1~meV and thus mainly
originate from dimers or clusters of spin islands close to each
other. Influenced by the nearby spin islands, the magnetization
profile thus deviates from the one of an isolated spin island.

In conclusion, by combining thermodynamic and
high-resolution inelastic neutron scattering techniques with numerical simulations, we were
able to quantitatively study the properties of a system of strongly interacting
spin islands formed in a strong-leg Heisenberg spin ladder depleted
with non-magnetic impurities. An effective description in the
language of these objects explains and quantitatively
reproduces the measured data. As long as the spin-ladder medium
remains quantum disordered at $H<H_c$, this effective description
is faithful and, in principle, would allow not only to describe
the thermodynamic response but even to access dynamical quantities
such as time-dependent correlation functions \cite{Unpub}. Beyond
$H_c$, however, the effective description breaks down due to the
quantum phase transition to the TLSL
state~\cite{giamarchibook1}, rendering the spin-ladder correlation
functions fundamentally different. To the authors
knowledge, the problem of non-magnetic impurities and the fate of
spin islands in the TLSL phase of a spin ladder or their influence
on the magnetically ordered state
\cite{giamarchi2008Np1,Schmidiger2012p1} has not been addressed
experimentally or theoretically. We hope that our
study stimulates further efforts and experiments towards
understanding the impurity-induced physics in such systems.

This work is partially supported by the Swiss National Science
Foundation, Division II. We would like to thank Dr. S. M\"uhlbauer
(FRM II, Technische Universit\"at M\"unchen) for his involvement at
the early stages of this project and Dr. S. Gvasaliya (ETH Z\"urich)
for assistance with the magnetic measurements. Simulations were performed
on the Brutus cluster (ETH Z\"urich).

% ----------------------------------------------------------------------------------------

%

% -----------------------------------------------------------

\pagebreak
\clearpage
\onecolumngrid

\begin{center}
\textbf{\large Supplementary material: \\ Emergent interacting spin islands in a depleted strong-leg Heisenberg ladder}
\end{center}
\vspace{0.35in}

\twocolumngrid

\setcounter{equation}{0}
\setcounter{figure}{0}
\setcounter{table}{0}
\setcounter{page}{1}
\makeatletter

% ----------------------------------------------------------------------------------------

\section{(A) Experimental details}

\subsection{Measurement details}

Single crystals of \DIMPYz\ were grown from solution by the temperature
gradient method \cite{Yankova_PhilMag_2012_ReviewXtalsA}, according to the same procedure as for the 
original material \DIMPY~\cite{Shapiro2007p1A} but with replacement of the
relevant amount $z$ of CuBr$_2$ by ZnBr$_2$. For the magnetic
measurements single crystals with typical masses of 15~mg were used. The
measurements were carried out with the help of a standard Quantum
Design Magnetic Properties Measurement System (MPMS-XL SQUID
magnetometer). For measurements below 1.7~K, the $^{3}$He cryostat
insert for MPMS (iQuantum iHelium3) was employed. 
For all the samples, the magnetic field
was applied along the $a$ axis of the structure.

For the synthesis of single crystals for neutron scattering, 
all the hydrogen-containing chemicals were replaced by the deuterated
analogues. Similar to our studies on the clean material 
\cite{Schmidiger2011p1A,Schmidiger2012p1A,Schmidiger2013p1A,Schmidiger2013p2A,PovarovSchmidiger_PRB_2015_DimpyScalingA},
the measured sample was composed of three co-aligned
crystals of (C$_7$D$_{10}$N)$_2$Cu$_{0.96}$Zn$_{0.04}$Br$_4$ with
typical masses of 800~mg each. They were fixed on a similar aluminum sample holder
as previously used (see e.~g.~Fig.~2 of Ref.~\cite{Yankova_PhilMag_2012_ReviewXtalsA}), 
with the $b$ direction being vertical.

\subsection{Spin island scattering in reciprocal space}

Hereby, we would like to provide an additional 
figure, showing the part of 
reciprocal space accessed in the IN5 neutron 
experiment as well as the 1D spin island
scattering.

\begin{figure}[h!t]
\centering
\includegraphics[width=0.85\columnwidth]{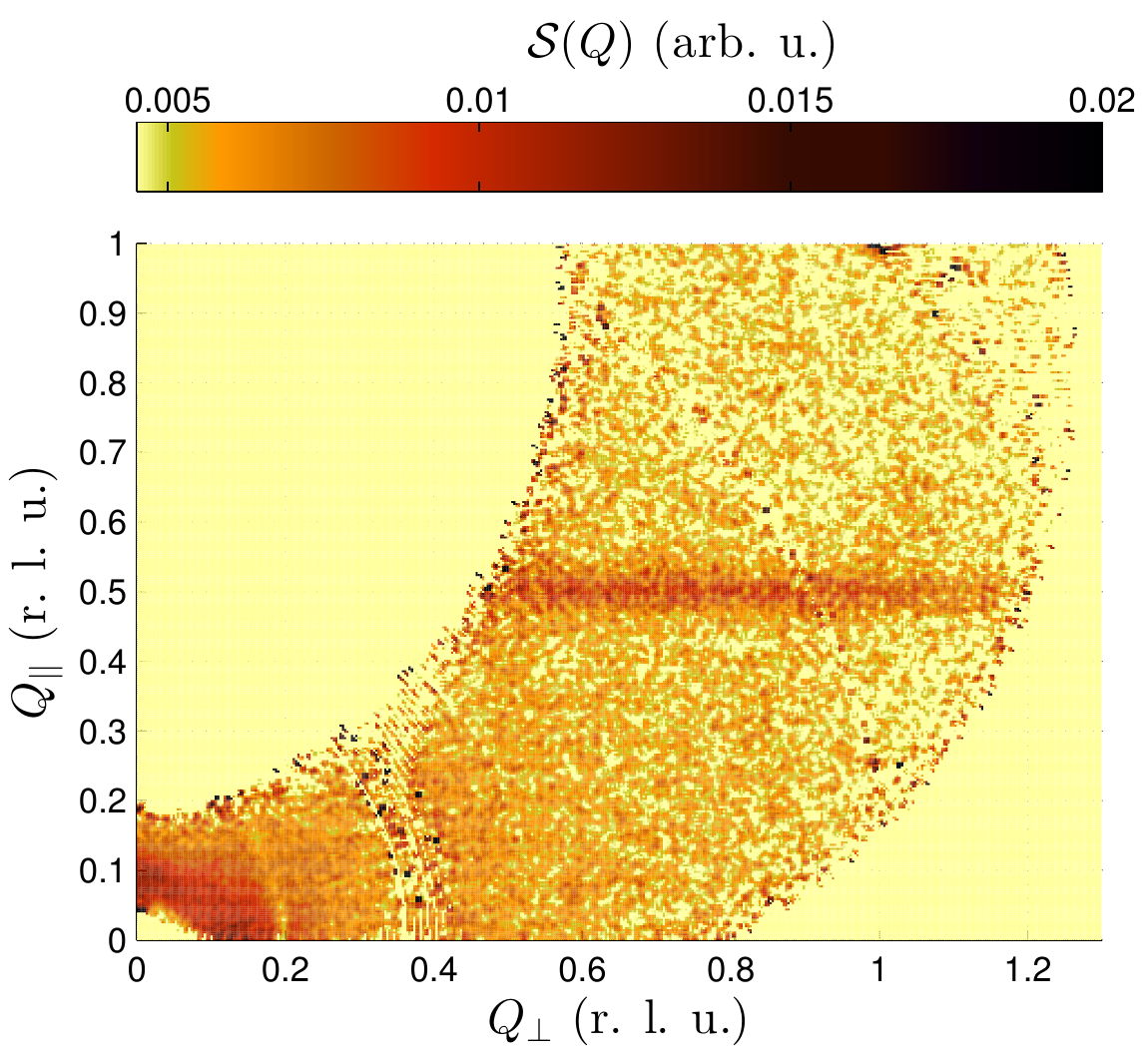}
\caption{Neutron intensity of \DIMPYdz~with $z=0.04$, 
as measured with $E_i = 1.3$~meV and $T=75$~mK. Uncorrected data is shown as a function of 
$Q_{||}$ (along the ladder leg) and the perpendicular in-plane direction. Data
was integrated in the out-of-plane direction $Q_k$ ($-1\,\mathrm{r.l.u.}\le Q_k \le 1\,\mathrm{r.l.u.}$) and in the 
energy range $0.1\,\mathrm{meV}\le\hbar\omega\le 0.3\,\mathrm{meV}$.} \label{fig:neutron}
\end{figure}

Data in Fig.~\ref{fig:neutron} (measured with the IN5 instrument \cite{Ollivier2011p1A}, 
with $E_i = 1.3$~meV and $T=75$~mK) was integrated along the 
out-of-plane direction $-1\,\mathrm{r.l.u.}\le Q_k \le 1\,\mathrm{r.l.u.}$ and in the energy
range $0.1\,\mathrm{meV}\le\hbar\omega\le 0.3\,\mathrm{meV}$.
The spin island scattering is readily observed as a stripe of 
intensity, located at $Q_{||}=0.5$~r.l.u..

\section{(B) Numerical details}

In this section, further details about the numerical 
simulations are provided. We follow closely the 
previous work \cite{Mikeska_DopingA,Lavarelo2013p1A}, studying
the general $J_{||} = J_\perp$ coupling case.

\subsection{Magnetization profile}

\begin{figure}[h!t]
\centering
\includegraphics[width=1\columnwidth,trim={1.5cm 1cm 12cm 10cm},clip]{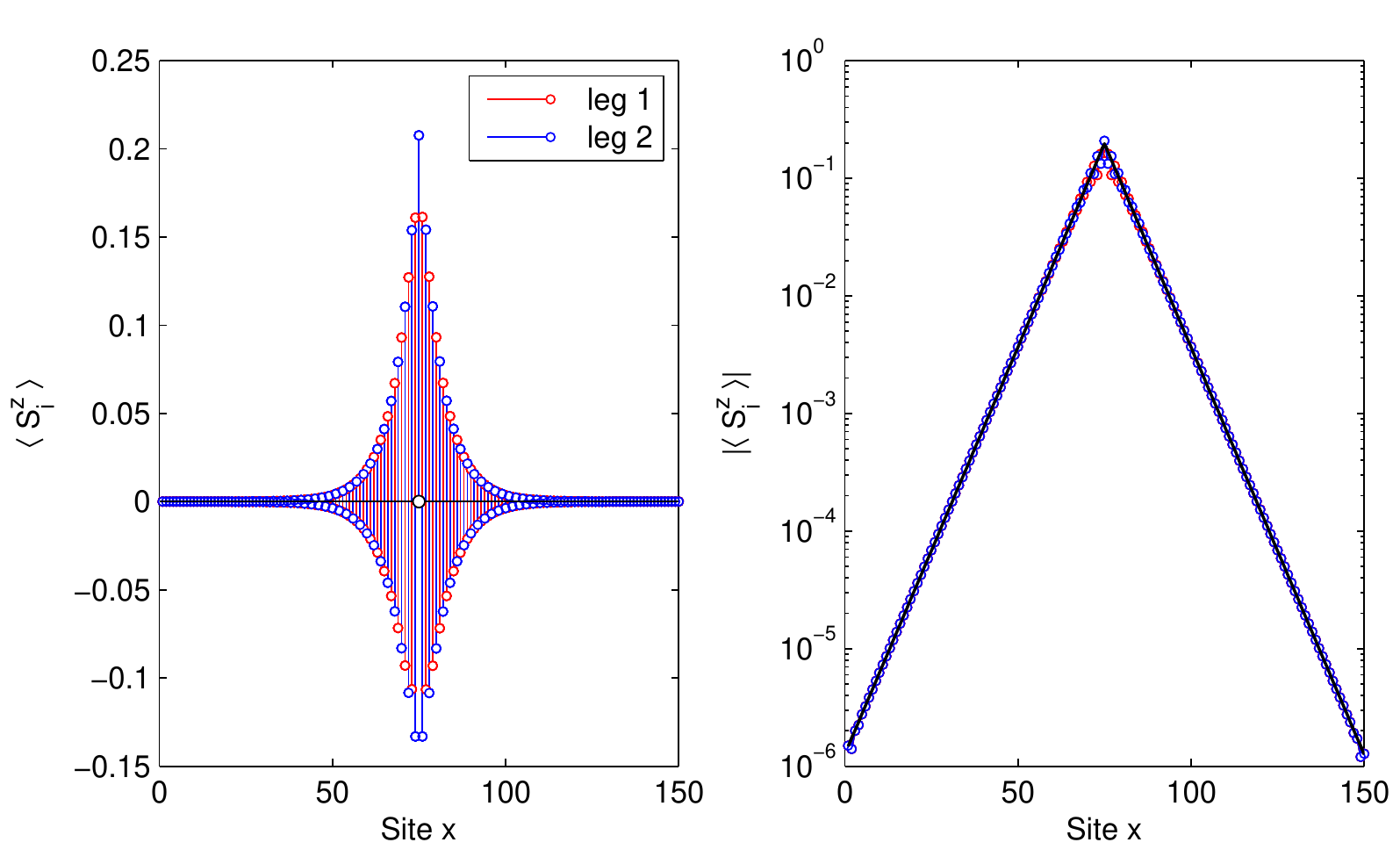}
\caption{Local magnetization $\langle S_i^z \rangle$ of a spin ladder with 
a single non-magnetic impurity in the center as calculated with the DMRG method
($J_{||}=1.42$~meV and $J_\perp = 0.82$~meV). Linear (left panel) and logarithmic (right panel)
visualization, together with the exponential fit (solid line).}\label{fig:shape}
\end{figure}

The shape of the emergent magnetization profile around 
an impurity was studied by placing a single $S=0$
site in the center of a ladder with $2\times150$ spins
and by calculating the local magnetization $\langle S_i^z \rangle$
of the ground state with $M_S = 1/2$. Fig.~\ref{fig:shape} illustrates
the obtained local magnetization $\langle S_i^z \rangle$ 
using the Density Matrix Renormalization Group (DMRG) 
algorithm from the ALPS libraries \cite{ALPSv2p0A} ($J_{||}=1.42$~meV 
and $J_\perp = 0.82$~meV, $350$ DMRG states kept). 
The local magnetization follows an exponential law without further 
corrections as expected \cite{Mikeska_DopingA,Lavarelo2013p1A}. By fitting 
$\log|\langle S_i^z \rangle|$ with a linear function, the 
exponential decay length was determined to be 
$\xi = 6.27(1)$.

The spin island form factor $F(Q_{||},Q_\perp)$ is obtained by Fourier 
transformation, 
\begin{align}
F(Q_{||},Q_\perp) = \sum_{n=-\infty}^\infty \sum_{m=0}^1 S_{n,m}\,\mathrm{e}^{\mathrm{i}2\pi nQ_{||}}\,\mathrm{e}^{\mathrm{i}m2\pi Q_\perp},
\end{align}
and with $S_{n,m}\approx S\,(-1)^{n+m}\,\mathrm{e}^{-|n|/\xi}$ and
$S_{0,0} = 0$ at the impurity site, the latter
simplifies to
\begin{align}
F(Q_\parallel,Q_\perp) = \frac{(1-\mathrm{e}^{-\mathrm{i}2\pi Q_\perp})\sinh(\xi^{-1})}{\cos(2\pi Q_\parallel) + \cosh(\xi^{-1})} - 1.
\end{align}

\subsection{Effective interactions}

If two impurities are introduced, two spin islands emerge, 
represented by effective spins ${\bf S}_{\bf I}$ 
at the impurity sites, with $S_{\bf I}=1/2$ \cite{Lavarelo2013p1A}. Their 
interaction depends on the distance along the leg ($L_x$) and rung $(L_y)$
and due to their effective interaction, they pair up to a dimer with
a $S=0$ singlet and a $S=1$ triplet state at energies well below 
the gap energy of the ladder system \cite{Mikeska_DopingA,Lavarelo2013p1A}. 
Depending on whether $L_x+L_y$ is odd or even, the effective 
interaction is ferro- or antiferromagnetic (FM or AFM).
The mutual effective interactions were determined based on Matrix Product State (MPS) \cite{Dolfi2014p1A} 
calculations of ladder systems with $2\times 150$ sites 
and two non-magnetic impurities placed at a predefined distance
$L_x$ along the leg and either in 'cis' or 'trans'
configuration (impurities on the same leg or different legs \cite{Mikeska_DopingA}). 
The two impurities were centered 
around the middle of the ladder. In order to minimize boundary effects, 
the distance to the ladder ends is larger than 4-5 correlation lengths. 
Following the approach described in Refs.~\cite{Mikeska_DopingA,Lavarelo2013p1A}, the two lowest
energies in the symmetry sectors 
with total spin projection of $M_S=0$ and $M_S=1$ were 
calculated. If spin-island interactions are AFM, the lowest
state corresponds to a singlet ($S=0$, $M_S=0$) while the first excited state is 
a triplet ($S=1$, $M_S=0,\pm1$) and vice versa for FM interactions. The energy difference
between the lowest two states in the $M_S=0$ sector thus 
corresponds to $J^\mathrm{eff}({\bf L})$ while the sign is
determined from the lowest energy state of the $M_S=1$ sector, 
as discussed in Refs.~\cite{Mikeska_DopingA,Lavarelo2013p1A}.

Calculations were performed with the matrix MPS 
algorithm from the ALPS package \cite{Dolfi2014p1A} (250 states kept). As 
an independent check-up, the results published in table II
of Ref.~\cite{Mikeska_DopingA} were successfully reproduced first. In the latter, a similar 
numerical study was performed for the  $J_\mathrm{leg}/J_\mathrm{rung}=1$
coupling case, close to the DIMPY coupling ratio of  $1.73$.

\begin{figure}[h!t]
\centering
\includegraphics[width=0.85\columnwidth,trim={2.5cm 1cm 14.5cm 10cm},clip]{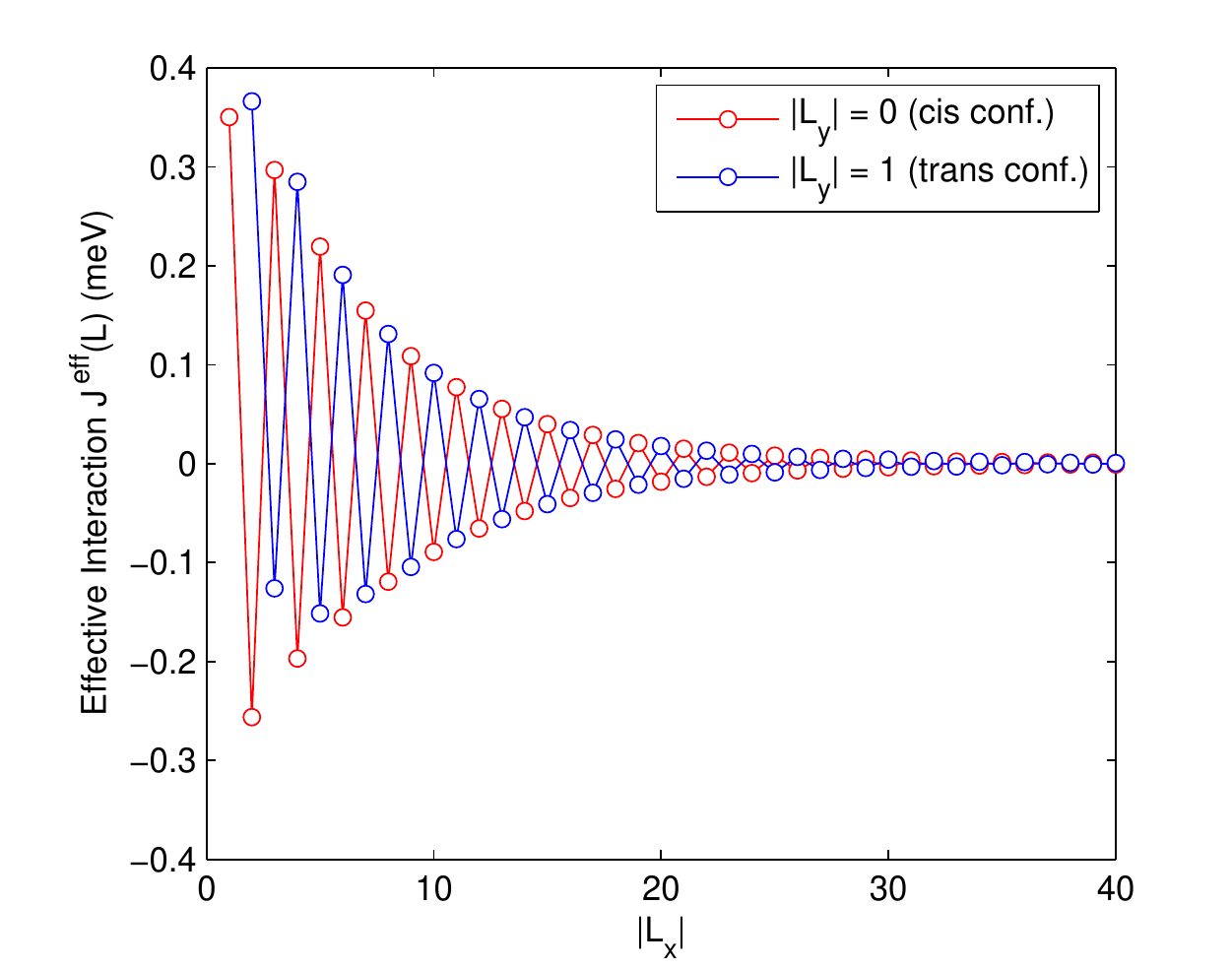}
\caption{Effective interactions $J^\mathrm{eff}({\bf L})$ obtained based
on DMRG calculations of two non-magnetic sites placed in the center of a spin ladder
with a mutual distance ${\bf L} = (L_x,L_y)$, as described in the text. Red and blue 
symbols correspond to non-magnetic sites at the same or different legs (with $L_y=0$
and $|L_y|=1$ respectively).} \label{fig:int}
\end{figure}

The determined effective interactions are shown in Fig.~\ref{fig:int}. As previously noted in 
Refs.~\cite{Mikeska_DopingA,Lavarelo2013p1A}, they follow
\begin{align}
J^\mathrm{eff}({\bf L}) = J_0\,(-1)^{L_x+L_y+1}\,\mathrm{e}^{-|L_x|/\xi}
\end{align}
where $L_x \in \mathbb Z$ and $L_y \in \{0,1\}$. The parameters $J_0$ and $\xi$ were determined to be $J_0=0.441(7)\,\mathrm{meV}$ and $\xi = 6.28(3)$ unit cells for the DIMPY coupling case, in agreement with 
the trend shown in Fig.~2 and Fig.~5 of Ref.~\cite{Lavarelo2013p1A}.

\begin{figure*}[h!t]
\centering
\includegraphics[,width=1\textwidth]{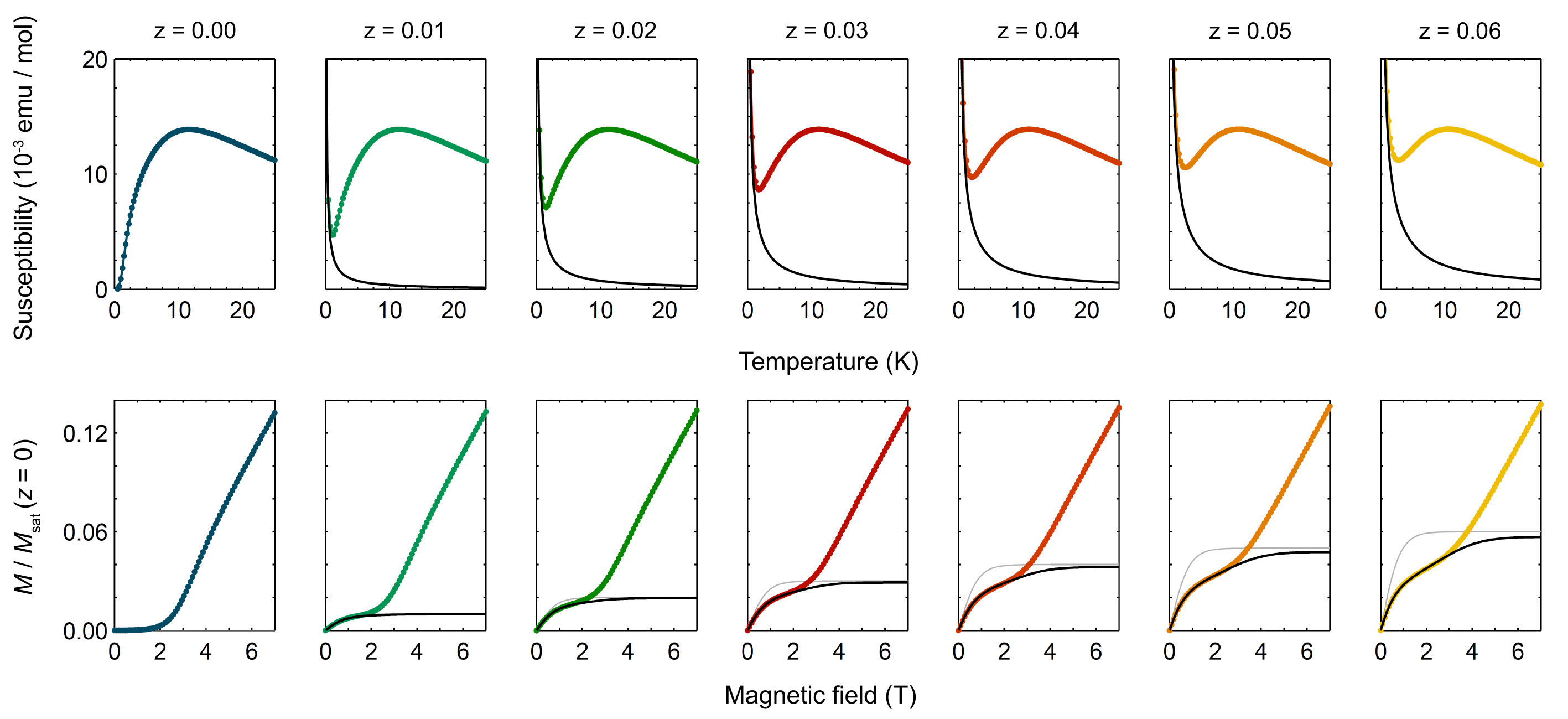}
\caption{Summary of numerical calculations of zero-field susceptibility $\chi(T)$
(upper panel) and field-dependent magnetization $M(T,H)$ at $T=600$~mK (lower panel), performed
for the coupling case of the depleted Heisenberg spin ladder \DIMPYz with $z\le 0.06$. Magnetization 
is normalized with respect to the magnetization of the pure $z=0$ model and thus corresponds 
to $2\langle S_i^z\rangle$, averaged over all magnetic and non-magnetic sites. Coloured 
symbols: QMC calculations of the full Hamiltonian with $N=2Lz$ non-magnetic sites and $L=500$ rungs, 
averaged over 300 random configurations. Black lines: Exact diagonalization of the low-energy effective 
model, as described in the text, with $N=12$ spin islands ${\bf S}_{\bf I}$ and 5000 random configurations. 
Gray line: Brillouin response assuming that each non-magnetic impurity is accompanied by a spin 
island and without interactions between different spin islands. } \label{fig:comparison}
\end{figure*}

\subsection{Thermodynamics: QMC calculations}

Since the original Heisenberg spin ladder Hamiltonian 
is non-frustrated, the thermodynamic properties are
accessible with numerical Quantum Monte Carlo (QMC)
calculations. The temperature-dependent susceptibility $\chi(T)$ and the
magnetization $M(T,H)$ were calculated using the QMC algorithms provided 
by the ALPS package \cite{ALPSv2p0A}. 
Calculations were performed for $J_{||} = 1.42$~meV and 
$J_\perp = 0.82$~meV, on depleted ladders with $2\times500=1000$ 
sites and a fixed number $N=z\cdot 1000$ of randomly placed non-magnetic sites. 
Similar to Ref.~\cite{Lavarelo2013p1A}, all thermodynamic quantities were averaged over 300
random configurations (total computation time: 2 months, parallelized to 
32-64 cores on the Brutus Cluster at ETH Z\"urich). The susceptibility was calculated 
using the 'loop' algorithm, for $0.5\,\mathrm{K} \le T \le 300\,\mathrm{K}$ 
(Fig.~\ref{fig:comparison}, coloured symbols in upper panel). 
The field-dependent magnetization was calculated as a function of 
magnetic field, for $0\,\mathrm{T}  \le H \le 8\,\mathrm{T}$, 
for $T=0.6$~K (Fig.~\ref{fig:comparison}, coloured symbols in lower panel). For the latter, 
the Stochastic Series Expansion (SSE) 
algorithm with directed loops \cite{SSESandvikp1A,ALPSv2p0A} was applied, similar to 
Ref.~\cite{Lavarelo2013p1A}.

\subsection{Thermodynamics: Effective model}

The low-energy model is described by effective $S=1/2$ 
spin operators at the site of impurities describing
the spin islands, interacting through \cite{Lavarelo2013p1A}

\begin{align}
\mathcal{H}_\mathrm{Sp.Isl.} = \sum_{{\bf I},{\bf J}} J^\mathrm{eff}({\bf I}-{\bf J}) \; {\bf S}_{{\bf I}} \cdot {\bf S}_{\bf J} - g\mu_\mathrm{B}H \sum_{\bf I} S_{\bf I}^z. 
\label{eq:HamEff}
\end{align}

Interactions $J^\mathrm{eff}({\bf L})$ depend on the distances ${\bf L}=(L_x,L_y)$,
as determined in the 'Effective interactions' subsection.

\begin{figure}[h!t]
\centering
\includegraphics[width=1\columnwidth]{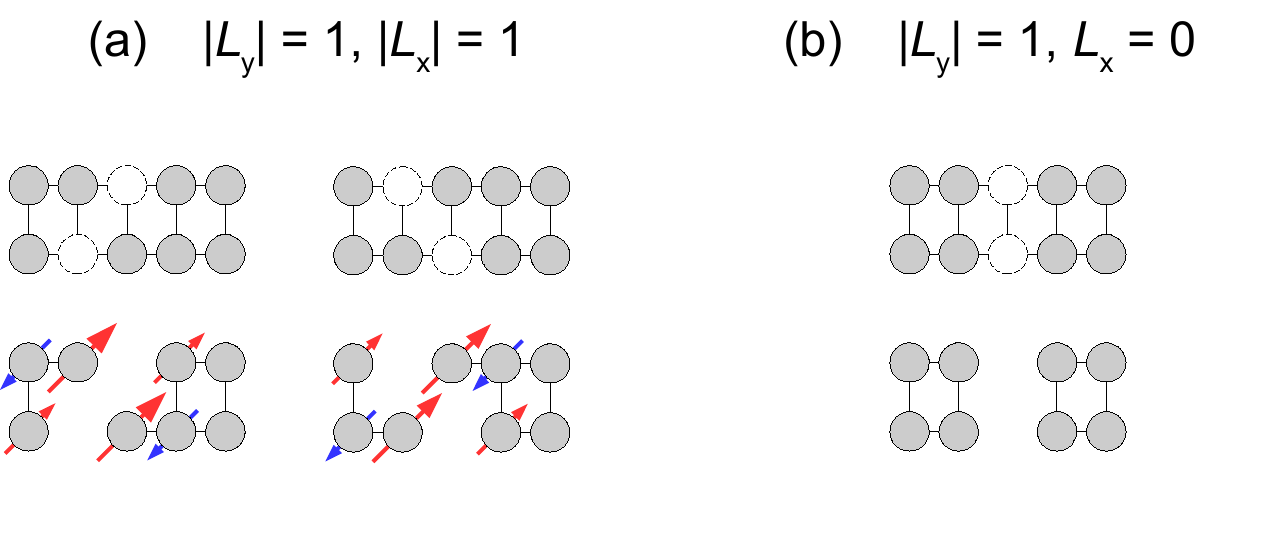}
\caption{Two special cases considered. (a) Impurities on adjacent 
rungs in trans position lead to fragmentation and release two impurities, one 
at each end. (b) Two impurities occupying the same rung lead to fragmentation 
but release no spin island at all.} \label{fig:specialcases}
\end{figure}

A small number of effective spins ${\bf S}_{\bf I}$ was randomly
placed on $N$ non-equal positions of a ladder with  $L=N/2z$ rungs. 
From this random arrangement, the mutual distances ${\bf L} = {\bf I}-{\bf J}$
and the corresponding interactions $J^\mathrm{eff}({\bf L})$ 
were determined. Before setting up the Hamiltonian, two special cases were
considered (Fig.~\ref{fig:specialcases}).

First, impurities on adjacent rungs 
in trans position ($|L_x|=1$ and $|L_y|=1$) split the system
into two non-interacting segments, releasing a 
spin island on both sides. Second, impurities on the same rung 
 ($L_x=0$ and $|L_y|=1$) cut the ladder into to non-interacting 
fragments as well. However, in contrast, the latter release no 
additional spin island.  Based on these requirements, the 
Hamiltonian was set up according to Eq.~(\ref{eq:HamEff})
and by exactly diagonalizing (ED) the latter, 
both eigenvalues $E_\mu$ with spin projection quantum 
number $M_{S,\mu}$ and 
eigenvectors $V_\mu$ were obtained.
From the former, thermodynamic quantities such as 
the magnetization and the susceptibility were calculated,
\begin{align}
M_\mathrm{r}(T,H) &= \frac{2z}{N}\,\langle M \rangle\notag\\
\chi_\mathrm{mol}(T,H) &= \frac{z N_\mathrm{A} (g \mu_B)^2}{N\,k_\mathrm{B}T} \Bigl(\langle M^2\rangle - \langle M \rangle^2\Bigr) 
\end{align} 
with $-1 \le M_\mathrm{r}(T,H) \le 1$, $h=g\mu_\mathrm{B}H$ and
\begin{align}
\mathcal{Z} &= \sum\limits_{\mu} \exp\biggl[-\frac{E_\mu -h\,M_{S,\mu}}{k_\mathrm{B}T}\biggr] \notag\\
\langle M^\alpha  \rangle &= \frac{1}{\mathcal{Z}}\, \sum\limits_{\mu} M_{S,\mu}^\alpha\,\exp\biggl[-\frac{E_\mu - h\,M_{S,\mu}}{k_\mathrm{B}T}\biggr].
\end{align}
The results were averaged over many repetitions $R$ of randomly chosen configurations
of $N$ impurities on a ladder with $L=N/2z$ rungs.

The susceptibility and field-dependent magnetization ($T=0.6\,\mathrm{K}$) are shown in Fig.~\ref{fig:comparison} as
black lines, for different $z$. Calculations were performed for a fixed number of $N=12$ impurities randomly
placed on ladders with $1200$, $600$, $400$, $300$, $240$ and $200$ sites for $z=0.01$, $0.02$, 
$0.03$, $0.04$, $0.05$ and $0.06$ respectively. The result was averaged over $R=5000$ random
configurations. By performing calculations for different number of impurities $2\le N \le 12$, we estimate
that despite the small number of impurities, the mean deviation of the thermodynamic quantities 
from the large-$N$ limit is less than 2\% for $z=0.06$ and even less for $z<0.06$. 
Calculations according to the effective model are similar to the ones published in Ref.~\cite{Lavarelo2013p1A} for the general
$J_{||}=J_\perp$ case (with $N=10$ effective spins and $R=1000$ repetitions), except for the additional special 
cases considered in this study (Fig.~\ref{fig:specialcases}). Notably and within the range of applicability,
the agreement between the ED calculations according to the effective model and the QMC calculation of the 
full depleted ladder Hamiltonian is excellent, 
for all concentrations considered and both for the susceptibility and magnetization.

We note that the ED treatment of the effective model provides access to the eigenvectors 
$V_\mu$ as well, thus enabling to calculate dynamical properties such as local correlations
$\mathcal{S}(\omega)$, measurable with inelastic neutron scattering \cite{Unpub}.

\subsection{Comparison to free spin model}

Finally, it is instructive to compare the QMC and ED numerical results in Fig.~\ref{fig:comparison}
to the simplified model based on the naive assumption that each non-magnetic ion is 
accompanied by a spin island and that the latter are non interacting, thus
leading to a Brillouin response (gray lines in Fig.~\ref{fig:comparison}) 
\begin{align}
\frac{M(T,H)}{M_\mathrm{sat}(z=0)} = z \, \tanh\biggl[\frac{g\mu_\mathrm{B}H}{2k_\mathrm{B}T}\biggr]. \label{eq:fs}
\end{align}
At very low concentration, this model becomes a faithful description 
since spin islands are well separated on average (mean distance $\overline{L}_x \gg \xi$)
and thus non-interacting. However, at $z \gtrsim 0.02$, the free spin model
becomes inappropriate in two ways. First, interactions between mutual spin islands
become progressively more important and deviations from the Brillouin response 
apparent. Second, the probability of impurities occupying the same rung 
increases as well. In the latter case (Fig.~\ref{fig:specialcases}b), no spin island is 
created at all. The probability that, in a system with impurity
concentration $z$, an impurity shares a rung with another impurity
is $p = z/2$ \cite{Lavarelo2013p1A}. Since two impurities are needed to form such
a non-magnetic rung pair and since both do not release a spin island, 
the saturated magnetic moment of the effective model is $z(1-z)$ 
instead of $z$ as in Eq.~(\ref{eq:fs}).

% ---------------------------------------------------------------------------------

%

\end{document}